# Resistive Switching in Aqueous Nanopores by Shock Electrodeposition

*Ji-Hyung Han[1,2,a], Ramachandran Muralidhar[3], Rainer Waser[4,5,6,b] and Martin Z. Bazant[1,2,b,c*]*

[1] Department of Chemical Engineering and [2] Department of Mathematics, Massachusetts Institute of Technology, Cambridge, MA 02139, USA * Email: bazant@mit.edu

[3] IBM Thomas J. Watson Research Center, Yorktown Heights, NY 1059

[4] Institut für Werkstoffe der Elektrotechnik II, RWTH Aachen University, Germany
[5] Peter Grünberg Institut 7, Forschungszentrum Jülich GmbH, Jülich, Germany
[6] JARA – Fundamentals of Future Information Technology

Present addresses: [a]Marine Energy Convergence & Integration Laboratory, Korea Institute of Energy Research (KIER), Jeju-do, 695-971 Korea; [b] Department of Materials Science and Engineering and [c] SUNCAT Center for Interfacial Science and Catalysis, Stanford University, Stanford, CA 94305

**Abstract:** Solid-state programmable metallization cells have attracted considerable attention as memristive elements for Redox-based Resistive Random Access Memory (ReRAM) for low-power and low-voltage applications. In principle, liquid-state metallization cells could offer the same advantages for aqueous systems, such as biomedical lab-on-a-chip devices, but robust resistive switching has not yet been achieved in liquid electrolytes, where electrodeposition is notoriously unstable to the formation of fractal dendrites. Here, the recently discovered physics of shock electrodeposition are harnessed to stabilize aqueous copper growth in polycarbonate nanopores, whose surfaces are modified with charged polymers. Stable bipolar resistive switching is demonstrated for 500 cycles with <10s retention times, prior to any optimization of the geometry or materials.

In recent years, memristive elements[1] based on internal redox-reactions have attracted intense interest for applications such as non-volatile Random Access Memory.[2-4] Resistive switching in solid-state ultra-thin films is widely viewed as a potential replacement for flash memory in devices requiring lower power, lower voltage, and higher programming speed. A popular class of these Redox-based Resistive Random Access Memories (ReRAM) is based on the reversible dielectric breakdown of transition metal oxide (TMO) films, in which oxygen vacancies (or transition metal interstitials) are moved and the valence change of the metal cations leads to the formation of conducting filaments.[4,5] This class is called Valence Change Memories (VCM), also known as OxRRAM. Another important class relies on an electrochemically active electrode material such as Ag or Cu and ultra-thin solid electrolyte or insulator films. This class is called electrochemical metallization memories (ECM), also known as conductive-bridge memories (CBRAM) or programmable metallization cells (PMCs).[2,3] In these systems, cation redox reactions lead to the growth and dissolution of nanoscale metal dendrites that reversibly short-circuit the electrodes to create two memory states with very different resistances.[2,3,32]

In contrast to solid-state devices, resistive switching in liquids has received much less attention, in spite of its potential to open new and different applications, e.g. for control, logic, or memory in microfluidic devices. Indeed, liquid-state ECM would appear to be impossible because rapid electrodeposition in bulk liquids becomes unstable to the growth of fragile, fractal deposits, which cannot be grown and dissolved reversibly.[6-8] To our knowledge, the only previous attempt at resistive switching in an aqueous system involved adding a drop of water in a gap between microelectrodes to induce the local formation and dissolution of Ag dendrites[9], albeit without demonstrating the long-term stability of ON/OFF cycles.

In this Letter, we report stable, reversible resistive switching in an aqueous system. The required control of metal electrodeposition[10,11] is achieved by exploiting the new physics of over-limiting current (faster than diffusion)[10,12-14] and deionization shock waves in charged nanochannels[10,15-18] and porous media[11,14,16,19-21]. In bulk liquid electrolytes, the morphological instability leading to dendritic growth results from diffusion limitation[22], but when the electrolyte is confined within negatively charged nanopores, surface transport of cations through the electric double layers can sustain over-limiting current and lead to the formation of stable deionization shock waves (jumps in ion concentrations and the electric field) that propagate against the applied current[12,18,20,21]. Deionization shocks have the general structure of a ``diffusive wave'', similar to the diffusion layer that precedes dendritic electrodeposition[22-24], but followed instead by a region of strong salt depletion in the nanopores, where large electric fields drive surface conduction and stabilize uniform metal growth[11,16]. In ordered nanopores with unmodified surfaces, metal nanowire growth is difficult to synchronize[25], but our group has showed that shock electrodeposition with charge-modified surfaces offers a powerful new means of controlling rapid metal growth, in both ordered[10] and disordered[11] nanopores. Interestingly, the field of solid-state ReRAM has just come to a similar realization[26], that ionic shock waves (in oxygen vacancies) also control the switching dynamics of memristive elements based on TMO thin films.

*Theory.* – We begin by explaining the fundamental connection between liquid-state and solid-state ion-concentration shock waves and their analogous applications in resistive switching devices. The mathematical structure of models in both fields is that of a nonlinear drift-diffusion equation,

$$\frac{\partial c}{\partial t} + v(c)\frac{\partial c}{\partial x} = D\frac{\partial^2 c}{\partial x^2}. \qquad (1)$$

In simple models of solid-state ReRAM, the local resistivity along a conducting path is proportional the oxygen vacancy concentration ($\rho \sim c$), and thus so are the local electric field ($E = \rho I$) and vacancy drift velocity ($v = \mu E$) at constant applied current ($I$). In that case ($v(c) \sim c$), the Nernst-Planck equation (1) reduces to Burgers equation, the simplest model of shock propagation by forward wave breaking ($vv' > 0$), which captures the essential nonlinearities of gas dynamics, water waves, and glaciers[27]. Physically, oxygen vacancies experience larger electric fields and drift faster in regions of higher concentration, so the crest of a concentration wave will overtake the trough until a steep gradient is stabilized by diffusion and propagates as a "shock" in the direction of the current.

In neutral binary electrolytes, concentration relaxation occurs by ambipolar diffusion without drift, but in charged electrolytes confined to micro/nanochannels[11,15,28] or "leaky membranes"[16,20,29], the oppositely charged internal surfaces act as dopants in semiconductors or transition metal sites in TMOs, leading to nonlinear drift by electromigration and electro-osmotic flow. The relative importance of surface-driven drift compared to bulk ambipolar diffusion grows as the electric field is amplified by decreasing salt concentration and enables over-limiting current, faster than diffusion[11,12,14]. For negatively charged surfaces, the effective drift velocity scales as $v(c) \sim -\rho_s I/(c - \rho_s)$, where $\rho_s < 0$ is the surface charge per volume, since excess cations in the double layers of concentration $-\rho_s$ drift in the local electric field ($E = \rho I$), which scales with the total resistivity ($\rho \sim (c - \rho_s)^{-1}$).[16,20] In this case, as in traffic flow[30], concentration waves break backwards, since crests move slower than troughs ($vv' < 0$). Physically, regions of salt depletion have greater resistivity and amplified electric fields, which drive further depletion by electromigration through the double layers. Electro-osmotic flows further enhance salt depletion via electro-hydrodynamic dispersion, which has thus far resisted a simple description by homogenized equations such as Eq. (1).[12-14,18,28,29]

Shock electrodeposition occurs when electrodeposition is preceded and regulated by a deionization shock wave. Since modeling is quite challenging with diffuse charge, electro-convection and moving boundaries, shock electrodeposition has only been studied experimentally to date[10,11]. In anodic aluminum oxide (AAO) with parallel 300nm pores and negatively charged surface coatings, shock electrodeposits grow uniformly along the surfaces at high currents, fed by surface conduction of cations through thin double layers[10]. In membrane materials with smaller, randomly intersecting pores, such as cellulose nitrate (200-300nm) and polyethylene (50nm), shock electrodeposition leads to macroscopically flat deposits, which can be reversibly cycled through the material.[11]

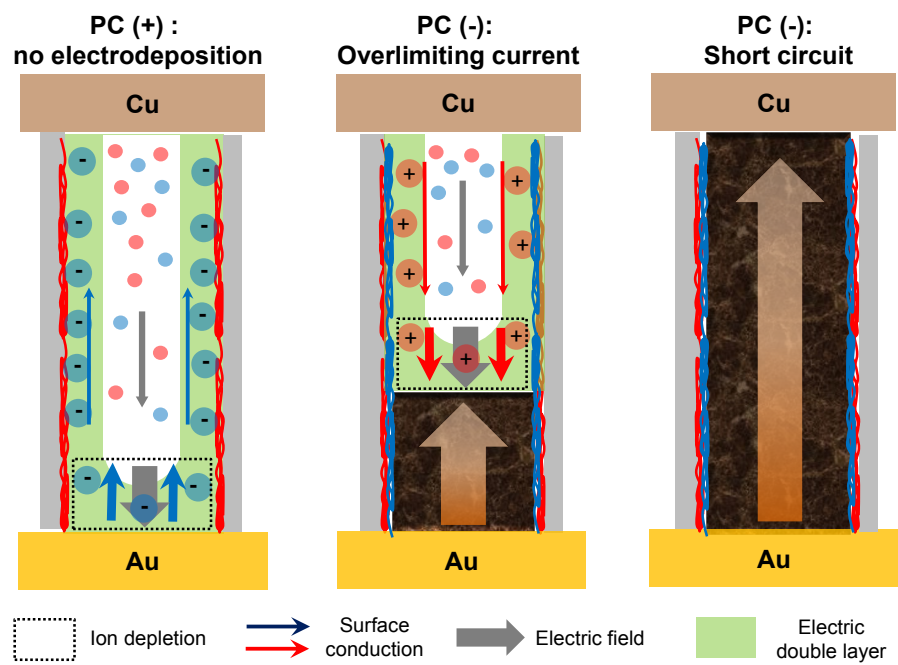

**Figure 1.** Cross-sectional sketch of deionization shocks forming ahead of metal electrodeposition by surface conduction in charged nanopores at high rates, as demonstrated in Ref. [10] for 300nm AAO nanopores with thin double layers and applied here to resistive switching in smaller 50nm PC nanopores with thick double layers in the ion depletion zone.

The observed stability of electrodeposition/dissolution cycles in these cases may be attributable to double layer overlap in the depleted region behind the shock wave, which leads to more uniform excess conductivity and thus more stable shock propagation, according to theory.[16] In order to observe deionization shocks, however, the surface charge must be small compared to the initial bulk salt concentration,[20] corresponding to thin double layers. Therefore, we predict that the ideal nanopore radius for programmable metallization by shock electrodeposition should lie in between the Debye screening lengths of the concentrated and depleted solutions, e.g. 10-300nm for 1mM-1µM, so that the double layers are thin ahead of the shock and thick behind it, as sketched in Fig. 1.

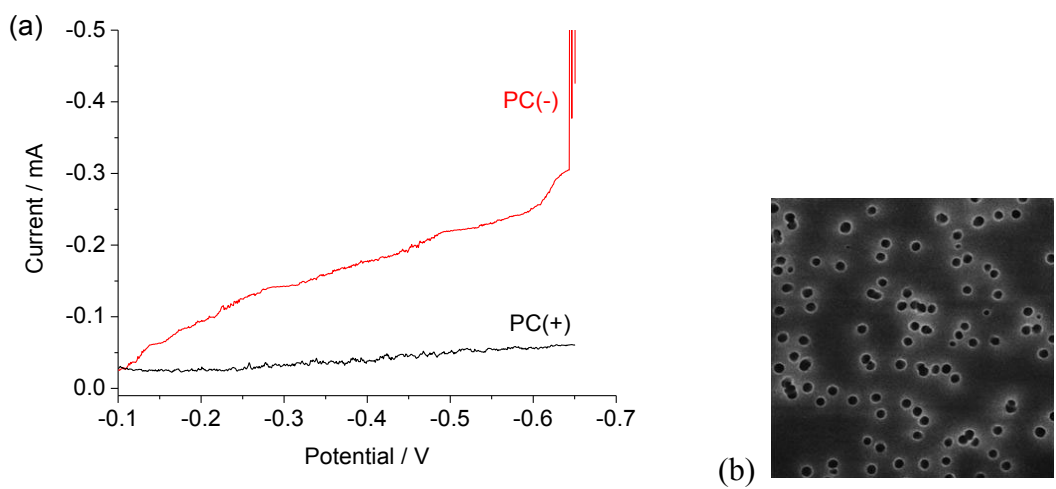

**Figure 2.** (a) Linear sweep voltammetry (LSV) of PC(+) and PC(-) membranes of exposed area 1 cm² between Cu disk electrode and Au wafer electrode in 1 mM $CuSO_4$ at 2 mV/s. The outer solution is water. The potential of the Au/Si working electrode is measured relative to the Cu counter-electrode, so that Cu electrodeposition on the working electrode (negative reduction current) occurs at negative voltages. (b) SEM image of a transverse cross section of a PC membrane, showing randomly dispersed, non-intersecting pores of 50nm mean diameter.

*Experiment.* – Our prototype for aqueous resistive switching utilizes polycarbonate (PC) membranes with ordered nano-sized cylindrical parallel pores (50 nm in diameter, 6 μm in length) whose surface charge is modified by layer-by-layer deposition of polyelectrolyte layers of alternating charge. Polydiallyldimethylammonium chloride (pDADMAC) is directly deposited on a bare PC membrane to make PC(+) with positive surface charge. A negatively charged PC(-) membrane is then obtained by coating negative polyelectrolytes (poly(styrenesulfonate), pSS) on the pDADMAC-coated PC(+) membrane. The PC membranes are soaked in 1 mM $CuSO_4$ solution, and an ECM cell is constructed, as in our previous experiments.[10] Briefly, the PC membrane is clamped between Cu disk electrode (13 mm diameter) and Au coated Si wafer (1 cm x 1.5 cm) under constant pressure. The cell voltage is reported as the potential the Au coated Si wafer working electrode, relative to that of the Cu counter-electrode. At potentials near Cu deposition/dissolution ($E^0$=0.16V), the Au substrate of the working electrode is electrochemical inert ($E^0$=1.50V). In order to prevent the evaporation of the electrolyte solution inside PC membrane, the electrochemical cell is immersed in a beaker containing distilled water (the "outer solution"). When 1 mM $CuSO_4$ is used as an outer solution, the net $Cu^{2+}$ ion flux into nanopores (balanced by dissolution at other locations of copper electrodes) can cause some pores to be permanently filled with Cu metal during resistive switching test. The use of distilled water as the outer solution allows us to achieve reversible switching characteristics over large numbers of ON/OFF cycles. It should be noted that there is some leakage of $Cu^{2+}$ ions flowing out of the nanopores because the PC membrane is not perfectly sealed against electrodes, especially at the metal surfaces evolve during electrodeposition/dissolution cycles. As small gaps appear between the PC membrane and the electrodes, it is expected that the concentration of the $CuSO_4$ "inner solution" within the nanopores undergoing reversible metallization will gradually decrease over time during cycling experiments.

The current-voltage curves (Fig. 2a) show the expected dependence on the surface charge of the porous medium during over-limiting current with metal electrodeposition.[10,11] The PC(-) exhibits a gradual increase in current and very sharp current jump at around -0.65 V, while current of PC(+) is relatively low and constant without current jump. The effect of surface conduction on metal electrodeposition rate is illustrated in Figure 1b. Surface conduction of cations in PC(-) provides a path for fast ion transport by electromigration (and to a lesser extent, electro-osmotic flow), driven by the large electric field in the ion depletion region, leading to short circuit current by the metallic contact of copper deposits with the opposite copper electrode. Since the depleted region behind the shock can reach micro-molar concentrations[14,21] with ~300nm Debye length, the double layers will strongly overlap, which should improve the stability of resistive switching by stabilizing the shock profile ahead of the growth, as noted above.

The morphologies of electrodeposits forming the short circuits were not characterized in this study, since it is difficult to isolate the relatively small number of nanopores filled with copper, but based on our previous imaging experiments on copper electrodeposits in anodized aluminum oxide (AAO) membranes with similar surface charge modifications,[10] it is likely that the short circuit structures consist of mostly space-filling copper nanorods, as well as some copper nanotubes coating the surfaces, depending on the variable local current density within each nanopore.[25] On the other hand, positive surface charge suppresses metal growth as a result of oppositely-directed surface conduction, and the copper deposits grown in PC(+) are typically very short and never approach the Cu counter-electrode, as in our previous work.[10]

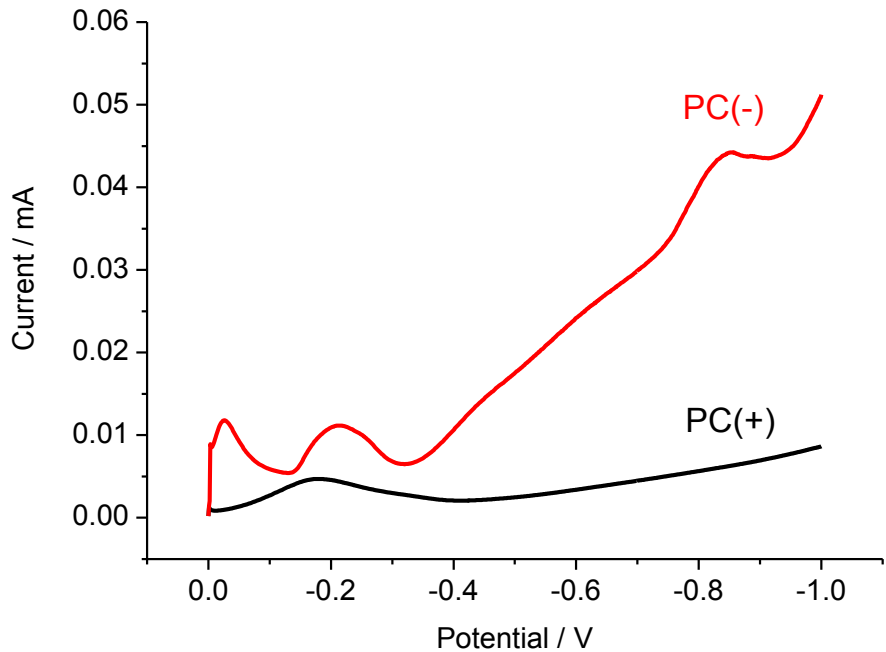

**Figure 3.** LSV of PC(+) and PC(-) membranes of exposed area 1 cm$^2$ between Cu disk electrode and Au wafer electrode in distilled water at 2 mV/s. The outer solution is also distilled water.

The current-voltage relations for PC(+) and PC(-) were also examined with distilled water as the inner solution (Figure 3). The current of PC(-) is much higher than that of PC(+), which provides evidence that growth of copper nanowires is driven by surface conduction of Cu$^{2+}$ ions through a dilute CuSO$_4$ solution within the nanopores, which originate from dissolution of the Cu anode. In distilled water, however, PC(-) does not show evidence of any short circuits, as clearly observed with the 1 mM CuSO$_4$ inner solution, because there are not enough Cu$^{2+}$ ions transported through the nanopores to produce significant growth of the Cu cathode.

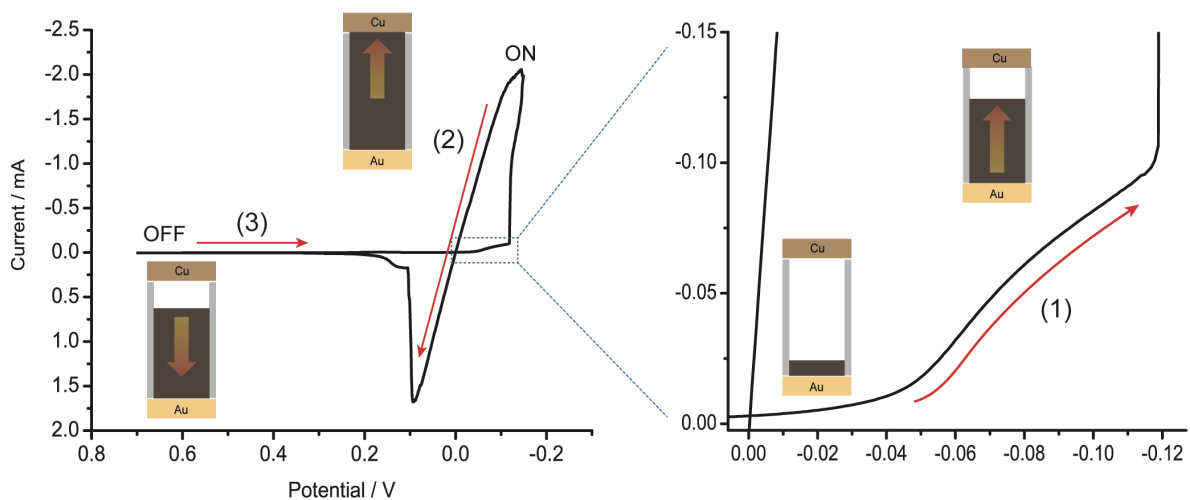

**Figure 4.** Resistive switching of by copper electrodeposition (brown) in the PC(-) membrane in 1 mM CuSO$_4$ at 1 mV/s with sketches of the stages of metallization within the charged nanopores.

After getting the first short circuit of PC(-) membrane from the LSV (Figure 2A), an oxidation potential (0.8 V) is applied at the Au electrode to detach Cu nanowires from the Cu electrode until the current is decreased to less than 0.05 mA. Repeated cycles of resistive switching are then measured using cyclic voltammetry (Figure 4). Some Cu nanowires start to grow through the nanopores during overlimiting current when negative potential is applied at Au electrode, followed by abrupt increase in current of approximately two orders of magnitude around -0.12 V, which implies that several Cu nanowires are in contact with Cu electrode to create a short circuit (ON state). The high conductivity of ON state begins to decline rapidly at 0.1 V, and the Cu filaments are detached from the Cu electrode around 0.2 V by substantial dissolution, which leads to OFF state. Further application of the oxidation potential dissolves some Cu dendrites grown at the Au electrode. The onset potential where short circuit occurs the first time (Figure 2A) is different from the corresponding onset potentials in subsequent ON/OFF cycles (Figure 4) because the bare Au metal substrate where Cu nanowires first start to grow is different from the surface covered with Cu residues in later cycles, which likely remain on the Au electrode after the OFF state. Indeed, it is well known that the overpotential for Cu electrodeposition on the same metal surface is much less than on a foreign substrate.[31]

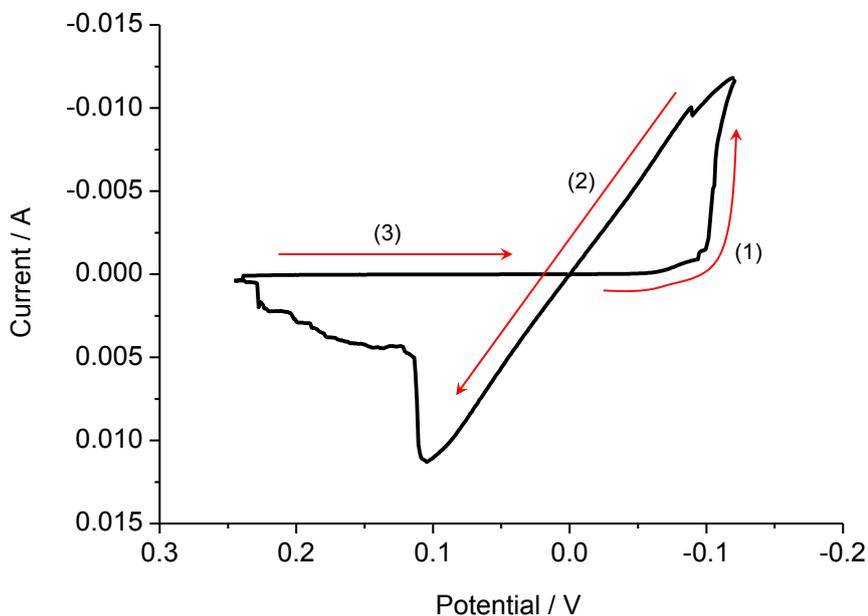

**Figure 5.** Resistive switching of copper in the AAO(-) membrane of exposed area 1 cm$^2$ between Cu disk electrode and Au wafer electrode in 10 mM CuSO$_4$ at 1 mV/s. The outer solution is water.

In order to test the effect of nanopore radius, we also studied resistive switching AAO(-) membranes with 300nm pores from our previous work[10], whose diameter is comparable to the Debye length in deionized water. In order to further reduce double layer overlap, we also used a larger initial salt concentration of 10 mM, which increases the depleted concentration at the same current[14,21]. Under these conditions, similar behavior of resistive switching was observed in AAO(-) as in PC(-), but the cycling capability is significantly degraded, consistent with the theoretical arguments above (Figure 5). The onset potential where short circuit occurs is around -0.1 V, which is almost the same with the PC membrane, but the current under the positive voltage before reaching the OFF state is quite different. The sharp decrease of high conductivity stops around 0.11 V, and then the current gradually declines, which suggests that the Cu nanowires in ON state do not detach from the Cu electrode at the same time.

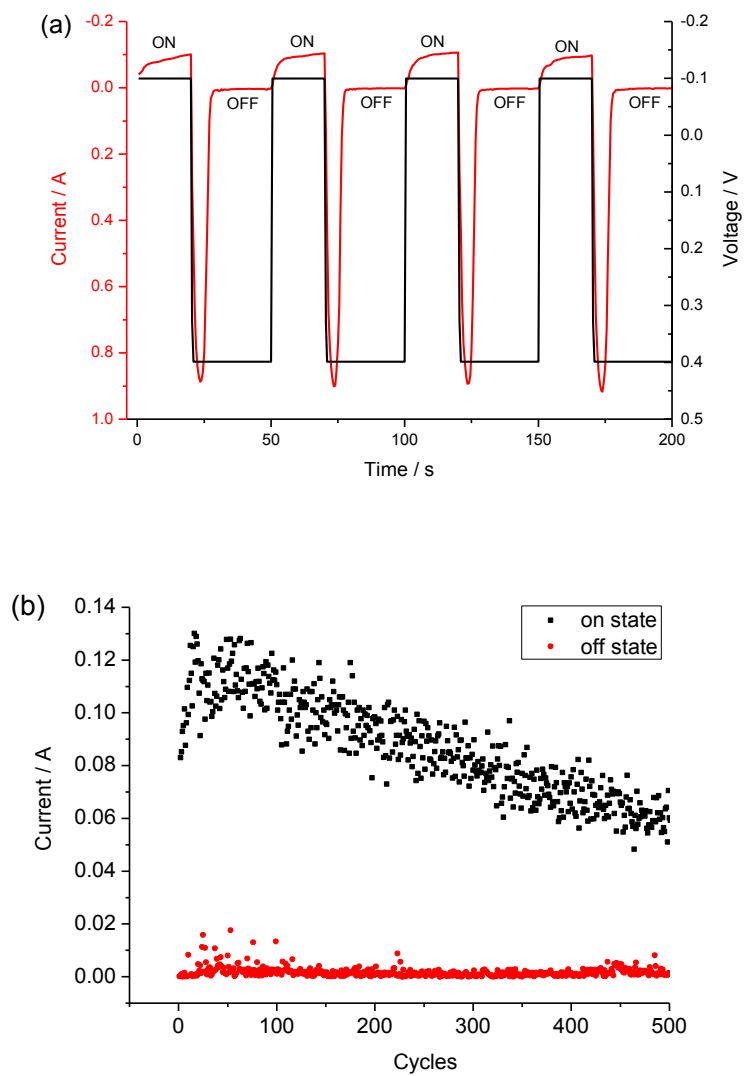

**Figure 6.** (a) Cycling test at negative pulses (ON state) of -0.1 V for 20 s followed by positive pulses (OFF state) of 0.4 V for 30 s. (b) Long-term ON/OFF cycling over 7 hours. Copper is electrodeposited and electrostripped in PC(-) in an inner solution of < 1 mM $CuSO_4$ and distilled water as the outer solution.

Cycling tests were performed with the most stable PC(-) membrane by applying repeated pulses of negative and positive voltages (Figure 6). The ON state current of - 0.1 mA was found to be very reproducible and stable (Figure 6a). A consistent retention time on the order of seconds (~7 s) is required to reach the OFF state after applying positive voltage. Long-term cycling between ON/OFF states for over 7 hours shows fairly good stability and reproducibility under the same experimental conditions (Figure 6b). Although the ON state current is gradually decreasing, the OFF state current remains almost constant, and robust resistive switching is observed throughout the experiments. Even after 500 cycles, the current still reliably jumps by two orders of magnitude between the ON and OFF states with a consistent response time < 10 s.

The slowly decaying ON state current may be attributable to the gradual exchange of ions between the inner and outer solutions, which could be controlled in future designs in order to extend the cycle life. We have already noted that a more concentrated outer solution tends to cause excess metallization as additional cations diffuse into the PC(-) nanopores, which leads to larger ON currents but also difficulty switching back to the OFF state,. In the present prototype with distilled water as the outer solution, cations slowly leak out from the nanopores, and the initial concentration of the inner solution (1 mM) becomes significantly reduced. As a result, the number of Cu nanowires in metallic contact with the Cu disk electrode in the ON state is decreased, leading to higher resistance and lower current. Nevertheless, we can still easily tell the difference between ON and OFF state currents after 500 cycles over 7 hours. The stability and reliability of ON/OFF switching could be improved by better sealing the negatively charged nanopores against the electrodes, surrounding them with positively charged polymer layers, adjusting the concentration of the outer solution, or other methods of controlling to cation exchange with the inner solution.

In summary, we have demonstrated robust resistive switching in an aqueous porous medium by copper metalization in negatively charged nanopores. The ECM-type switching process exploits the stability and control of metal electrodeposition afforded by deionization shock waves, which form during the passage of over-limiting current by electromigration through the electric double layers of the nanopores. After some optimization of the geometry, materials and protocols, it should be possible to develop individually accessible nanopores for aqueous ReRAM with subsecond response times, lasting for thousands of ON/OFF cycles. Applications of aqueous resistive switching include control in lab-on-a-chip devices for micro-total analysis of chemical samples, biomedical diaognostics or drug delivery, and related logic and memory.

This work was supported in part by a grant from the MIT Energy Initiative. MZB acknowledges partial support from the Global Climate and Energy Project at Stanford University and by the US Department of Energy, Basic Energy Sciences through the SUNCAT Center for Interface Science and Catalysis.